%% file: v750.tex
\documentclass[12pt, times,twocolumn ]{aastex701}

\usepackage{xspace}
\graphicspath{ {./}{figures/} }
\usepackage{wrapfig}

\newcommand{\mang}   {\,\mathrm{{\mbox{\AA}}}\xspace}
\newcommand{\kev}    {\,\mathrm{keV}}
\newcommand{\ks}     {\,\mathrm{ks}}
\newcommand{\scnd}     {\,\mathrm{s}}
\newcommand{\pc}     {\,\mathrm{pc}}
\newcommand{\mk}     {\,\mathrm{MK}}
\newcommand{\msun}   {\,M_\odot}

\newcommand{\chan}   {{\it Chandra}\xspace}
\newcommand{\hetg}   {{HETG}\xspace}
\newcommand{\xmm}    {{\it XMM-Newton}\xspace}

\newcommand{\nustar} {{\it NuSTAR}\xspace}

\newcounter{ion} \newcommand{\eli}[2]{\setcounter{ion}{#2}#1{~\sc\roman{ion}}}

\newcommand{\mone}    {^{-1}}
\newcommand{\mtwo}    {^{-2}}

\newcommand{\cmmtwo}  {\,\mathrm{cm\mtwo}}

\newcommand{\eflux}   {\,\mathrm{erg\,cm\mtwo\,s\mone}}
\newcommand{\kms}     {\,\mathrm{km\,s\mone}}
\newcommand{\lum}     {\,\mathrm{erg\,s\mone}}
\newcommand{\mdot}    {\msun\,\mathrm{yr}\mone}
\newcommand{\pflux}   {\,\mathrm{photon\,cm\mtwo\,s\mone}}

\newcommand{\vsfz}{ {V750\,Ara}\xspace}
\newcommand{\piaqr}{{$\pi\,$Aqr}\xspace}
\newcommand{\gcas}{{$\gamma\,$Cas}\xspace}
\newcommand{\gcastype}{{$\gamma\,$Cas-type}\xspace}

\newcommand{\mki}{
  Kavli Institute for Astrophysics and Space Research, 
  Massachusetts Institute of Technology, 77 Massachusetts Ave., 
  Cambridge, MA 02139, USA
}

\newcommand{\Change}[1]{{\bf #1}}
\renewcommand{\Change}[1]{#1}

\newcommand{\ChangeB}[1]{{\bf #1}}
\renewcommand{\ChangeB}[1]{#1}

\begin{document}

\title{{\it Chandra}/HETG and {\it NuSTAR} Observations of V750~Ara, a $\gamma\,$Cas-type Star}

\author[orcid=0000-0002-3860-6230,sname=Huenemoerder,gname=David]{David P.\ Huenemoerder}
\affiliation{\mki}
\email[show]{dph@mit.edu}

\author[orcid=0000-0003-2602-6703, sname=Gunderson, gname=Sean]{Sean J.\ Gunderson}
\affiliation{\mki}
\email{seang97@mit.edu} 


\author[orcid=0000-0002-7204-5502, sname=Ignace,  gname=Richard]{Richard Ignace}
 \affiliation{Department of Physics \& Astronomy, East Tennessee State University, Johnson City, TN 37614 USA}
 \email{ignace@etsu.edu}

\author[orcid=0000-0003-3298-7455, sname=Nichols, gname=Joy]{Joy~S.\ Nichols}
\affiliation{Harvard \& Smithsonian Center for Astrophysics, 60 Garden St., Cambridge, MA 02138, USA}
\email{jnichols@cfa.harvard.edu}

\author[orcid=0000-0002-6737-538X, sname=Pollock, gname=Andrew]{A.~M.~T. Pollock}
\affiliation{Department of Physics and Astronomy, University of Sheffield, Hounsfield Road, Sheffield S3 7RH, UK}
\email{A.M.Pollock@sheffield.ac.uk}

\author[orcid=0000-0002-1131-3059, gname=Pragati, sname=Pradhan]{Pragati Pradhan}
\affiliation{Department of Physics and Astronomy, Embry-Riddle Aeronautical University
3700 Willow Creek Road
Prescott, AZ, 86301, USA}
\email{pradhanp@erau.edu}

\author[gname=Norbert, sname=Schulz]{Norbert S.\ Schulz}
\affiliation{\mki}
\email{nss@space.mit.edu}

\author[orcid=0000-0003-1898-4223, gname=Dustin, sname=Swarm]{Dustin K. Swarm}
\affiliation{Department of Physics and Astronomy, University of Iowa, Iowa City, IA, USA}
\email{dustin-swarm@uiowa.edu}

\author[orcid=0000-0002-5967-5163, sname=Torrej\'{o}n, gname=Jos\'e]{Jos\'e M. Torrej\'{o}n}
\affiliation{Instituto Universitario de F\'{\i}sica Aplicada a las Ciencias y las Tecnlog\'{\i}as, Universidad de Alicante, E-03690 Alicante, Spain}
\email{tbs}

\begin{abstract}

We present a $197\ks$ \hetg and $95\ks$ \textit{NuSTAR} spectra of the
\gcastype object \vsfz.  The high-resolution X-ray spectra show that
the target is similar to other objects of this class.  Data are
interpreted under the assumption that the X-rays come from an
accreting white dwarf, and our analysis implies an accretion rate of
about $3\times10^{-11}\mdot$.  Emission lines are weak, and
predominantly \Change{ from hydrogen-like ions: \eli{Mg}{12},
  \eli{Si}{14}, and \eli{S}{16}.  H-like and He-like Fe are both
  present, but Fe~K fluorescence is weak, being significantly detected
  only in the \nustar spectrum, but was not obviously detected in the
  \hetg dispersed or zeroth-order spectra.}  The flux was variable above a level
expected by Poisson statistics.  There were no significant changes in
the spectral hardness, though we are limited by lack of soft signal
below $1\kev$.  Emission lines of Mg and Si were strong enough to
measure velocity offsets and widths which \Change{were found to be
  marginally inconsistent.  The H-like Mg line is consistent with
  instrumental broadening only, but shows a $300\kms$ blueshift.
  He-like Mg and H-like Si lines have no significant shift in velocity
  but are broadened by about $1000\kms$.  This suggests either
  different physical origins or velocity structure differing with
  plasma temperature.}

\end{abstract}


\section{Introduction}\label{sec:intro}

There are currently about 25 known stars in the \gcastype class
\citep{naze:motch:al:2020}.  They are late O- to early B-type (O9-B3)
emission line Oe and Be stars.  In the optical and UV, there are no
distinguishing features between the \gcastype stars and ``normal'' Oe/Be
stars \Change{\citep{smith:oliveira:al:2016, naze:motch:al:2020}}.  In
X-rays, however, there is a remarkable difference: the \gcastype
spectrum is very hard ($\sim10\kev$).  In embedded-wind-shock O- and
B-stars (that is, excluding colliding-wind systems and Be-X-ray
binaries with a neutron star), the temperatures are softer at
$0.5$--$1.0\kev$.  Both the normal stars and the \gcastype objects have
X-ray emissions consistent with thermal plasmas.  The \gcastype stars also
sometimes display strong Fe~K fluorescence emission, which is not a
feature of O- and B-star winds.  \gcastype \Change{ X-ray luminosities
  ($0.5$--$10\kev$) tend to be a bit higher, near $10^{32}\lum$
  \citep{lopesde:2006, naze:motch:2018, naze:motch:al:2020}.  Most Be
  stars, which can have later spectral types than the limit} of known
\gcastype stars, are much less luminous in X-rays
\citep{motch:oliveira:al:2015}.

The number of cases has grown from the eponymous member to enough
stars that they now comprise a class.  Recent searches for \gcastype{s}
among Be stars using \xmm has doubled their number
\citep{naze:motch:2018}.  Properties of \gcas itself were reviewed by
\cite{smith:oliveira:al:2016}, along with a synopsis of about a dozen
class members known then.  
\Change{They speculated on possible origins of the hard X-ray flux
  from magnetic reconnection between the star and disk, to accretion
  onto a compact companion.  \citet{postnov:al:2017} and
  \citet{tsujimoto:al:2018} also suggested that \gcastype stars'
  strong X-rays could be due to accretion onto neutron stars or white
  dwarf (WD) companions.  \citet{langer:baade:al:2020} investigated
  binary star evolution scenarios and found it probable that Be stars
  have a WD, neutron star, or black hole companion.  They suggested
  \gcastype stars are otherwise normal Be stars which have a
  helium-star companion whose low mass and low luminosity would make
  them hard to detect optically, but whose strong wind interactions
  with the Be star produce the observed high-temperature X-rays.  Many
  of the \gcastype stars are now confirmed to be binaries, probably
  with a low-mass \ChangeB{companion} ($0.6$--$1.0\msun$)
  \citep{naze:rauw:czesla:al:2022}, but none are known to have
  main-sequence companions \citep{bodensteiner:shenar:al:2020}. 
}

Given the fundamental distinguishing criterion of hard X-ray flux, and
the suggestion of the role of binarity and colliding winds, it is
important to obtain both broad band and high resolution X-ray spectra
of these stars.  The high resolution characterization of emission
lines is needed to resolve gas kinematics.  Broad-band spectra are
needed to constrain plasma temperatures reaching $\sim10\kev$, and
thereby the implied shock velocities under the hypothesis of colliding
winds, wind-disk interaction, or accretion.

Several \gcastype{s} have been observed at high resolution with \hetg:
\gcas \citep{oliveira:smith:al:2010, smith:cohen:al:2004}, HD~110432
(BZ~Cru) \citep{torrejon:schulz:al:2012}, HD~42054, and \piaqr
\citep{huenemoerder:pradhan:al:2024}.  They have line widths
$\lesssim1000\kms$.  For a stripped core He-star, wind speeds are
expected to be $\sim2000$--$3000\kms$ \citep{vink:2017}.  Hence widths
of $\sim1000\kms$ are compatible with this; the observed line 
broadening depends on details of wind-shock geometry, wind acceleration,
relative momentum flux of each wind, shock-cone cooling profile, and
viewing angle \citep{Henley:al:2003, rauw:al:2016}.  For comparison,
the embedded wind shock scenario in OB-stars produces temperatures of
$\sim0.3$--$0.8\kev$ ($3.5$--$9.0\mk$) \citep{cohen:parts:al:2021}. In
the single O4 supergiant, $\zeta\,$Puppis, the maximum temperature is
about $1.4\kev$ ($10\mk$), equivalent to $900\kms$ relative velocity
for a strong shock \citep{huenemoerder:ignace:al:2020}, in a wind with
terminal velocity of about $2000\kms$
\citep{sander:hamann:al:2017}. In contrast, the massive colliding wind
binary, WR~25, has line widths of about $2000\kms$, more comparable to
the terminal velocities when observed at an aspect favorable to
maximum shock cone line widths \citep{pradhan:huenemoerder:al:2021}.

Based on X-ray spectral and timing characteristics,
\citet{huenemoerder:pradhan:al:2024} suggested that \piaqr --- and
by implication other \gcastype stars --- host white dwarfs, which
are possibly magnetic, and hence the X-ray emission from \gcastype stars
is similar to that of cataclysmic binaries, a scenario previously
proposed by \citet{tsujimoto:al:2023} and \citet{klement:rivinius:al:2024}.

\vsfz (HD 157832) was discovered to be a \gcastype star with \xmm by
\citet{lopes:motch:2011} by virtue of a very hard spectrum
($kT\sim11\kev$) for its B1.5Ve spectral type.  It was recently
determined to be a $95\,$day period binary by
\citet{wang:gies:al:2023} through low-amplitude ($6\kms$) periodic
radial velocity variations in Balmer $\alpha$ and $\beta$ emission
lines.  They estimated the mass of a companion to be about
$0.75\msun$.  They did not detect any spectral signature of the
companion and tentatively concluded that the companion is too faint to
be a sub-dwarf stripped-envelope O-star (sdO).
\Change{From He lines in the optical, \citet{lopes:motch:2011} found a
  projected rotational velocity of about $220\kms$.}  We list stellar
parameters for \vsfz in Table~\ref{tbl:starparam}.

\input{v750_tbl_starparam}

\section{Observations and Modeling}\label{sec:obs}

We obtained nearly $200\ks$ of exposure on \vsfz with the \hetg over a
period of 20 months, with one $95\ks$ contemporaneous \nustar
observation. All data were processed with standard programs for each
observatory (CIAO \citep{CIAO:2006} and {\tt 
nuproducts}\footnote{\url{https://heasarc.gsfc.nasa.gov/docs/software/lheasoft/help/nuproducts.html}}) appropriate to the 
times of observations.  An observation log is given in Table~\ref{tbl:obsinfo}.

\input{ v750_tbl_obs }
\subsection{Variability and Timing}

\begin{figure*}[ht]
  \centering\leavevmode
  \includegraphics*[width=0.55\linewidth, viewport=0 0 525 320]{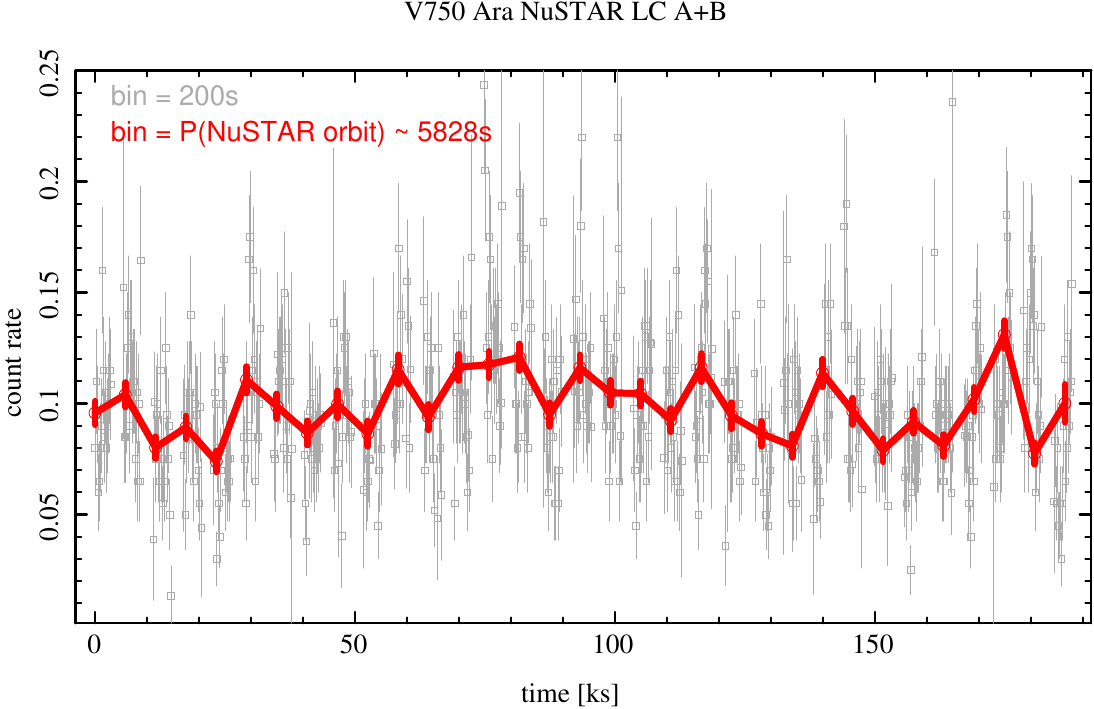}
  \includegraphics*[width=0.35\linewidth]{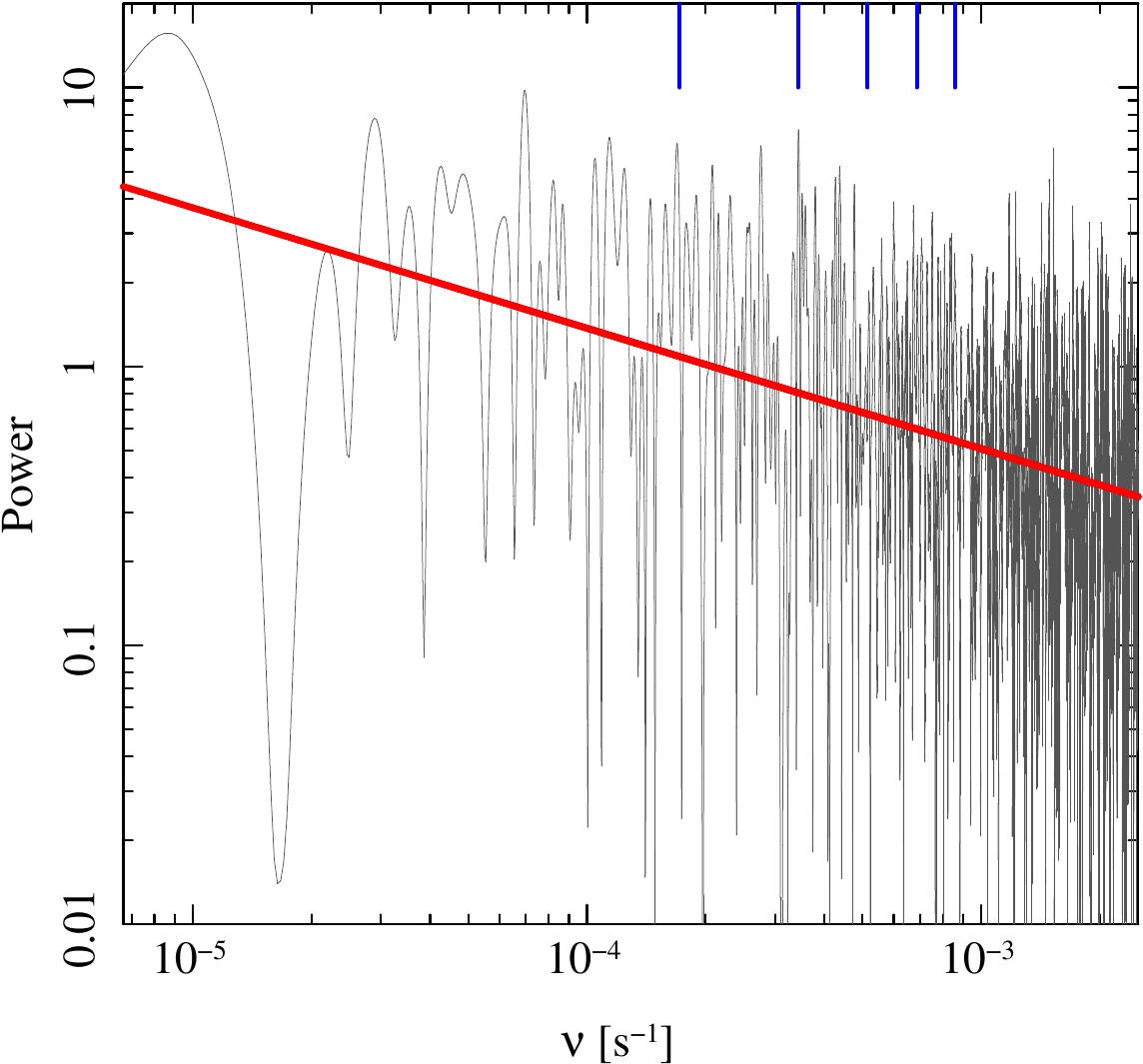}
  \includegraphics*[width=0.91\linewidth]{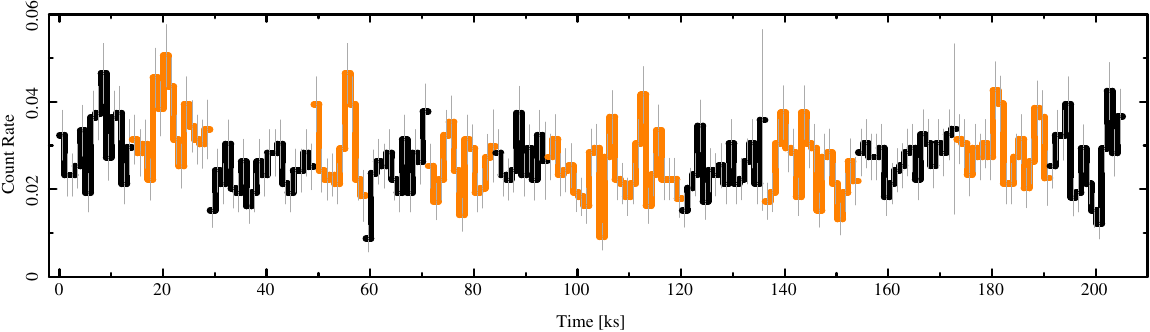}
  \caption{
    \nustar light curve (top left) in $200\,$s (gray) and
 $5828\,$s bins (red).  The larger bin size was chosen to match the
 period of the \nustar orbit.  On the top right we show the
 power-spectral-density of the light curve which shows a slope
 indicative of red-noise -- flickering at amplitudes larger than
 expected from Poisson noise. We did not have enough sensitivity to
 reach the Poisson noise floor.  \Change{Blue tic marks at the top indicate
 the \nustar orbital frequency and \ChangeB{harmonics}, none of which are evident
 in the transform.}  Bottom: the \hetg light curve of dispersed events
 in $1\ks$ bins, with the observations concatenated in time order in
 alternating colors.  The \nustar observation is between the two
 observations at about $95\ks$.
 \Change{Count rates are given in $\mathrm{s}\mone$.  The
 power-spectral-density is a unitless normalized probability.}
 } \label{fig:nslc}
\end{figure*}

The source had a relatively constant count rate during the
observations.  There was excess variance, above that expected for the
mean count rate, in both the \chan and \nustar observations.  This is
a well known phenomenon of \gcastype stars, and is referred to as
``flickering'' \citep{bruch:1995, bruch:2021}.  In
Figure~\ref{fig:nslc} we show the \nustar light curve for two bin
sizes, and the power-spectral-density distribution (PSD).
\Change{The PSD was computed with with the Lomb-Scargle algorithm
  \citep{scargle:1982, horne:baliunas:1986}, as implemented in
  ``isisscripts''\footnote{\url{http://www.sternwarte.uni-erlangen.de/isis/}},
  a method which is noted for robustly handling unevenly sampled data.
  }
The coarsely binned light curve (by $5828\scnd/$bin, the \nustar
orbital period) clearly shows semi-regular excursions which are
several times larger than the statistical uncertainties.  The PSD
shows a clear slope, typical of red noise.  To verify the reality of
this slope, we also formed the PSD after randomizing the order of
rates within the same time bins (hence preserving the window function)
and obtained a flat PSD.  The \chan data showed similar fluctuations
in its light curves (Figure~\ref{fig:nslc}, bottom panel).  We
detected no large excursions in the count rate, or in hardness ratios,
though the latter had its sensitivity limited by counting statistics
and the effective band pass.

\subsection{Spectral Analysis}

\begin{figure*}[ht]
  \centering\leavevmode
    \includegraphics*[width=0.95\linewidth]{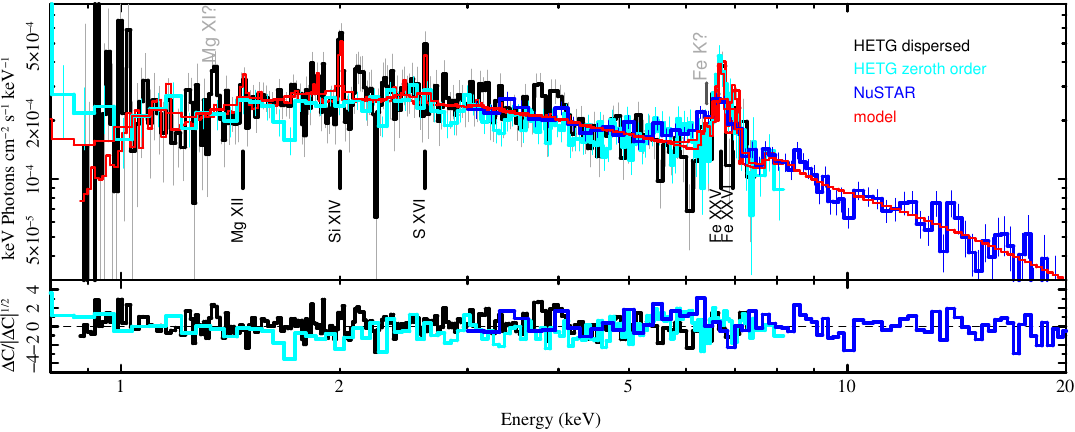}
 \caption{HETG
  spectrum of \vsfz in black, which is the merged HEG and MEG first
  orders, the HETG zeroth order in cyan, and the \nustar spectrum in
  blue (combined FPMA and FPMB).  In red is a model using \ChangeB{an absorbed
  cooling-flow model plus} a Gaussian for a tentative Fe~K
  fluorescence line.  Features with significant detections are marked
  below the spectrum.
  \Change{The bottom panel shows the residuals in the form of the
  delta-Cash statistic ($\Delta C$), appropriate for Poisson distributed counts,
  normalized by its square root absolute value ($|\Delta
  C|^\frac{1}{2}$), to make it analogous to the commonly used
  normalized $\Delta\chi$ statistic. Colors are as in the upper panel.
  }
} \label{fig:spec}
\end{figure*}

Since the count rates for both the \chan and \nustar observations
were essentially constant, we formed spectra over the entire interval
for each instrument.  The \chan count rate was also low enough in the
HETG zeroth order ($\sim0.03\,\mathrm{count\,s^{-1}}$) that CCD photon
pileup is not an issue, so we also include the zeroth-order spectra.
Figure~\ref{fig:spec} shows the flux-corrected spectrum for the three
cases.  No cross-calibration factors were applied.  Of particular note
is the paucity of emission lines.  \eli{Mg}{12}, \eli{Si}{14},
\eli{S}{16}, and \eli{Fe}{25} are significantly detected.  \eli{Mg}{11}
might be present, while \eli{Fe}{26} and Fe~K are probably present,
but relatively weak.  The overall shape is characteristic of a very
high temperature plasma, which is also consistent with the lack of
significant lower Z, lower temperature \ChangeB{ion states} with a narrow range of
formation temperature, such as the He-like lines, whereas H-like ions
have a long tail of emissivity to high temperatures.

For a spectral model, we used an absorbed ``cooling-flow'' model, used
in many astrophysical situations, in particular X-ray generation from
accretion shocks \citep{pandel:al:2005, tsujimoto:al:2018}.  
Our working hypothesis is that the X-ray emission
from \gcastype stars is likely from a WD companion to the
Be star.  A cooling flow is a plasma model in which the emission
measure distributions (EMD) relative weights as a function of
temperature are inversely proportional to the equilibrium plasma
cooling time at each temperature.  Additionally, the EMD range can be
clippped at a high and low temperature outside of which there is no emissivity, 
and the EMD can be modified by
a power-law factor, which represents deviations from an isobaric case.
In WD accreting systems, the normalization is proportional to the mass
accretion rate, and the maximum temperature is related to the WD mass.
Additional model parameters of possible use at high resolution are the
line width and velocity offset.  For both high and low resolution
spectra, the elemental abundances are also necessary parameters.
\citet{luna:raymond:al:2015} present a thorough overview of the
cooling flow physics and areas of application.   Figure~\ref{fig:spec}
shows our best fit model and residuals.  Model parameters are given in 
Table~\ref{tbl:plasmafit}.
\Change{The cooling flow model was implemented in the ISIS package
  \citep{houck2000} using the atomic database (AtomDB) of
  \citet{foster2012} for emssivities and wavelengths.}

\begin{figure}[ht]
  \centering\leavevmode
    \includegraphics*[width=0.90\columnwidth]{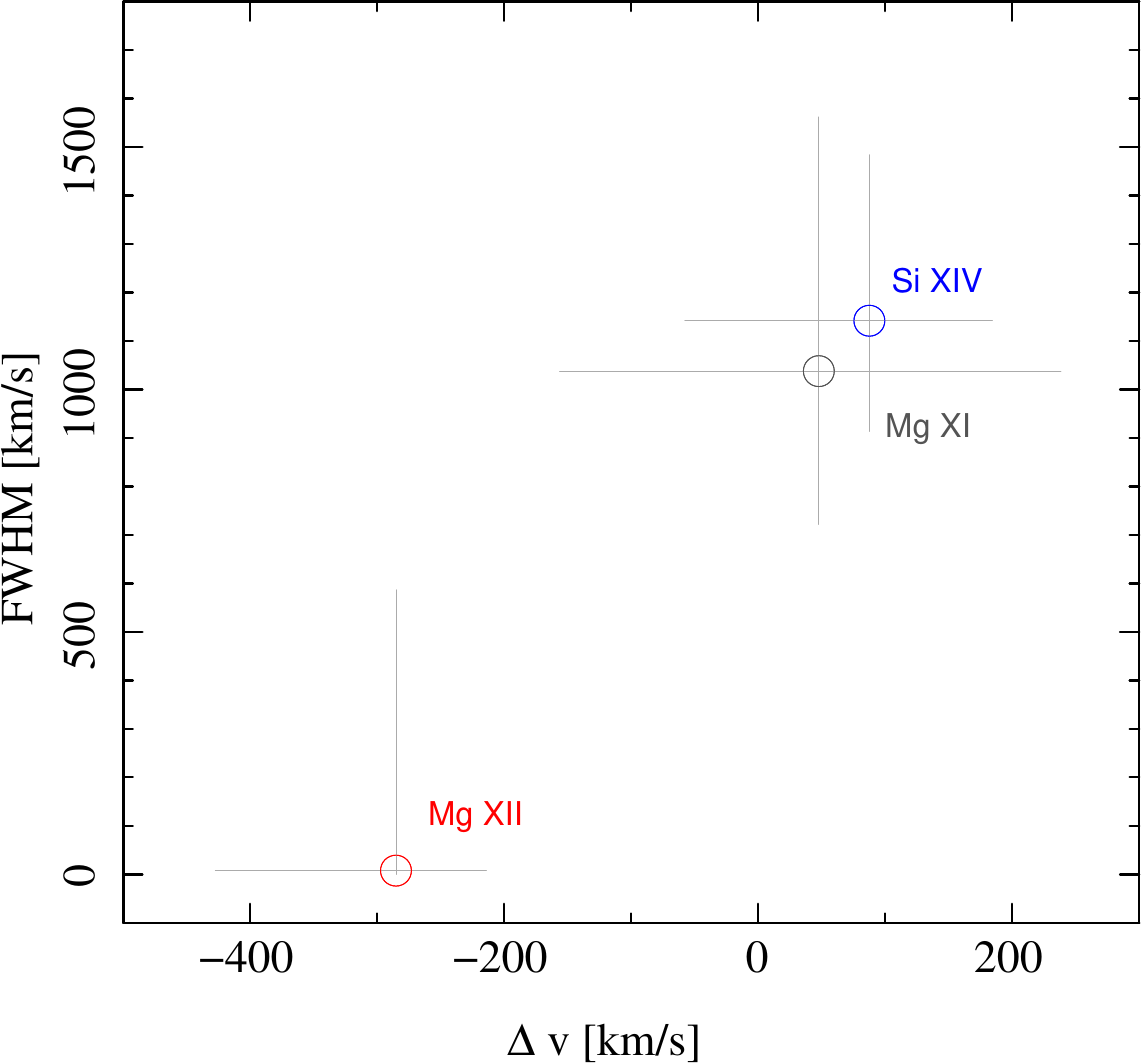}
 \caption{Line widths and offsets from Gaussian fits. Only features
 which provided meaningful constraints are shown.  Confidence limits
 are $1\sigma$. \Change{H-like feature velocities are in reference to
 the emissivity weighted mean of the doublet's wavelengths. The
 \eli{Mg}{11} He-like resonance, intercombination, and forbidden lines
 had their relative positions constrained and a common width
 parameter.  Widths are the intrinsic value required in addition to
 instrumental broadening.}
} \label{fig:linepar}
\end{figure}

We also fit features locally with Gaussian profiles folded through the
instrument response to investigate any dynamical effects on line
widths and centroids.  
\Change{
This provides some semi-empirical properties independent of global
spectral models, though we use the plasma model as a guide for
defining Gaussian components and initial parameters.  For instance,
the H-like lines of \eli{Mg}{12} and \eli{Si}{14} are doublets with a
2:1 emissivity ratio in collisionally ionized, low density plasmas.
Using Mg as an example, the separation of the components is about
$200\kms$, and the thermal width at maximum emissivity ($10\,$MK) is
about $140\kms$, mildly blending the components.  However, The HETG
instrumental resolution at $8.4\mang$ is about $700\kms$ (MEG) or
$360\kms$ (HEG) (FWHM of the instrumental profile), much larger than
the intrinsic width.  Furthermore, the global plasma model indicates
that there is additional broadening, which was treated as a turbulent
(Gaussian) component in addition to the thermal width, of about
$1000\kms$, just resolved by the HETG. Thus, single Gaussian models
are adequate to characterize the H-like emission lines.  Even so, at
high signal-to-noise, the single-Gaussian centroid of the H-like
doublet is determined by the emissivity-weighted mean position.
In referencing the centroid to velocity offsets, we have used
the weighted mean positions, though this difference is much smaller
than the statistical uncertainties in these observations.

For more crowded regions than those of \eli{Si}{14} and \eli{Mg}{12},
we included additional components, but imposed relative wavelength offsets
and used a common width parameter.  Such was necessary for the \eli{Mg}{11}
and the complicated $6\kev$ Fe regions.
There is also the possibility of unresolved blends, in particular from
a host of weak, unresolved dielectronic recombination satellite lines
which lie to the low-energy side of the resonance lines.  We used the
global plasma model and detailed emissivity data in the AtomDB
\citep{foster2012} to estimate their flux contribution to a spectral
region.  For \eli{Si}{14} and \eli{Mg}{12}, we find only about a $3\%$
contribution, but for \eli{Mg}{11}, \eli{Fe}{25} and \eli{Fe}{26}, the
contribution can be $10$--$15\%$.
Fits to the Fe region also included the HETG zeroth order and \nustar
spectra, along with the HETG first order spectrum.  Since the Fe~K
feature seemed present in the \nustar data but not in the HETG
observation, we have fit the region both jointly in all observations,
but also in the \nustar spectrum alone. 
}
The line fluxes are given in Table~\ref{tbl:plasmafit}.  Line widths
and offsets are shown in Figure~\ref{fig:linepar}.

%
\input{v750_tbl_fit_params}

\section{Discussion}\label{sec:disc}

The X-ray spectrum of \vsfz is largely as expected for a \gcastype
star, namely thermal, based on line emission, and very hard,
characteristic of very high plasma temperatures.  The \nustar spectrum
very strongly constrains the highest temperatures present to about
$260\mk$ ($22\kev$).  This hard component ($2$--$20\kev$) also
``flickers'', showing variance in excess of Poisson statistics
(Figure~\ref{fig:nslc}), a trait also known among \gcas stars, and common among cataclysmic variables 
\citep{bruch:2021}.

Under the assumption that the X-rays originate from Be-star disk
accretion onto a WD, the model normalization is proportional to the
mass accretion rate, and the maximum temperature is related to the WD
mass which controls the accreting plasma's velocity.  The
normalization (see Table~\ref{tbl:plasmafit}) implies $\dot M =
10^{-10}\mdot$, easily within mass expected to be available from a Be
star disk, but much less than typical of low-mass X-ray binary Roche
overflow rates \citep{patterson:raymond:1985a,
  patterson:raymond:1985b}.  This difference is reasonable, since the
reservoir of accreting material is not efficiently funneled onto a WD
from a Be star disk.  The value is also similar to that of \gcas
\citep[$3\times10^{-10}\mdot$;][]{gunderson:al:2025}, despite the
different system parameters.

For a WD mass estimate, we can use the relation between maximum
temperature, mass, and radius, in conjunction with a nominal WD
mass-radius relation \citep{ezuka:ishida:1999}, as was done for \gcas
by \citet{gunderson:al:2025}, to obtain $M_\mathrm{WD} =
0.9\pm0.1\msun$.  This is consistent with the estimate from the
single-lined spectroscopic binary solutions of
\citet{wang:gies:al:2023}, who found a mass of $0.8\msun$ if the
inclination were $i=60^\circ$; the mass would be larger for lower $i$.
Our values for \vsfz place it directly on the regression of
$T_\mathrm{max}$ vs. $M_\mathrm{WD}$ shown by \citet[][their Figure
  6]{tsujimoto:al:2018}.

The emission lines in \vsfz are quite weak, but we can use the
stronger features at the longest wavelengths available in the high
resolution spectrum to constrain flow velocities.  Only \eli{Si}{14},
\eli{Mg}{12}, and a tentative identification of \eli{Mg}{11} are
useful, and the width and offset values obtained from Gaussian fits
are shown in Figure~\ref{fig:linepar}.  Widths and offsets are
relatively small when compared to strong O-star stellar winds
\citep{pradhan:huenemoerder:al:2023} or to sdO winds which can have
velocities of $2000\kms$, though some are lower
\citep{mereghetti:al:2016, jeffery:hamann:2010}.  The \eli{Mg}{11}
feature is not actually predicted by our plasma model.  The tentative
identification has a low significance of about $2\sigma$.  It is our
lowest temperature feature, and if real, could be from a cooler
component of plasma emission, as was seen in \gcas by
\citet{gunderson:al:2025}.  Here we have moderate absorption and also
poor instrumental response to softer X-rays, so we cannot constrain
the cooler plasma with these data.

The $6-7\kev$ Fe-complex shows strong features, mostly due to
\eli{Fe}{25} and \eli{Fe}{26}; these are also indicative of the high
temperature plasma.  Furthermore, they have line-to-continuum ratios
requiring relative abundances of about $1.3$ times Solar
(Table~\ref{tbl:plasmafit}), in contrast to \gcas which is very
sub-Solar at a relative abundance of $0.33$ \citep{gunderson:al:2025,
  smith:cohen:al:2004}. The Fe~K fluorescent line is not present in
the HETG spectrum, though statistics are poor.  It does seem required
by the \nustar data.
\Change{
  Hence, we give two flux values in Table~\ref{tbl:plasmafit}, one for
  a joint fit of all data, and one for a fit to the \nustar spectrum alone.
}
If the \gcas model of \citet{gunderson:al:2025}
is scaled to the flux level and iron abundance of \vsfz, then an Fe~K
fluorescent line would easily be seen in the HETG spectrum with about
twice the flux as determined from our joint fit to the HETG dispersed
spectrum, zeroth order, and \nustar spectrum. There are two possible
reasons for this. First, the weakness could be indicative of system
geometry, which limits our view of fluorescent emission. Second, it
could be due to much less dense circumstellar disk material at the WD
site. Assuming a stellar radius of $5R_{\odot}$ for V750 Ara (B1.5Ve),
the white dwarf would orbit at $a_{\rm V750}\approx 40R_{*}$; in
contrast, \gcas companion orbits at $a_{\gamma}\approx 35R_{*}$. The
orbital radius ratio is then $a_{\rm V750} \approx 1.14
a_{\gamma}$. For a typical Be disk density profile, assuming similar
central densities and $\alpha=3.5$, this means $\rho_{\rm V750} \sim
0.6 \rho_{\gamma}$ at the site of the WD. The line emissivity depends
on the plasma density. Thus the line in V750 would be about two times
weaker than in \gcas. Furthermore, we have derived a WD mass similar
to that of \gcas. Yet the x-ray luminosity and implied mass acccretion
rate are a few times lower than \gcas's values, further reducing the
illumination of the reprocessing material. Since scaling the \gcas
line to the V750 flux predicts a strength only a few times greater
than what is observed, it is not surprising that we do not detect the
line here.

\Change{ The strength of the line in the \nustar spectrum, however, is
  more than twice the mean value fit jointly over all the spectra, and
  is thus actually consistent with the scaled value of \gcas.  This
  suggests that the fluorescent line is variable, even though the
  overall energy distribution of the \nustar spectrum agrees nicely
  with that from HETG.  }

In short, under this WD accretion hypothesis, these differences are
likely indicative of the different Be star disk conditions at the
locations of the WD, and differences in the Be star disks due to
stellar spectral type and rotation, which change the amount of
material available for accretion or the efficiency at which the WD can
accrete.

\section{Conclusions}\label{sec:conc}

We have obtained \chan/\hetg X-ray spectra of the \gcastype star, V750
Ara over a time span of almost 2 years, with one contemporaneous
\nustar observation.  We have interpreted the data in the context of
accretion onto a WD companion.  While the data do not prove that the
companion is a WD, they are consistent with that assumption.  The
important characteristics are:

\begin{itemize}

\item the spectrum is very hard, as born out by the \nustar observation;

\item spectral features are weak, except for \eli{Fe}{25} and \eli{Fe}{26}, also characteristic of high 
    temperatures;

  \item X-ray emission lines are slightly broadened, with intrinsic full-width-half-maximum of about $1000\kms$ 
    larger than instrumental, but not indicative of strong winds ($\gtrsim 2000\kms$), nor of magnetically confined  
    plasmas (unresolved);

  \item the X-ray flux displays flickering, a trait common in non-magnetic cataclysmic variables;

  \item the mean flux and hardness showed no large changes, despite the long coverage; no phase-dependent or 
    discrete absorption events were detected, as have been seen in
    other systems such as \gcas \citep{gunderson:al:2025} or \piaqr
    \citep{huenemoerder:pradhan:al:2024};

\Change{
  \item there may be variable Fe K fluorescence; its reduced mean
  strength relative to \eli{Fe}{25} and other \gcastype stars must be
  due to different accretion geometry; yet the \nustar data alone
  contradicts this, showing a well-detected feature, consistent with
  scaled \gcas flux.
}

   \item the mass we derive for the WD is about $0.9\msun$, consistent
   with the dynamical estimate of
     \citet{wang:gies:al:2023}, and consistent with the mass-temperature relation shown by 
     \citet{tsujimoto:al:2018}.

\end{itemize}

In summary, \vsfz, at high resolution, still looks like a \gcastype star and is consistent with the companion being 
a WD.

\bigskip

\begin{acknowledgments}

Support vfor SJG and DPH was provided the Smithsonian Astrophysical
Observatory (SAO) Chandra X-Ray Center (CXC) Award GO3-24004A, and by
NASA through the SAO contract SV3-73016 to MIT for Support of the CXC
and Science Instruments. CXC is operated by SAO for and on behalf of
NASA under contract NAS8-03060.  SJG and HMG also received support by
NASA through Chandra Award Number GO3-24005X issued by the CXC. JMT
acknowledges the grant ASFAE/2020/02 from Generalitat Valenciana.

This research has made use of ISIS functions (ISISscripts) provided by
ECAP/Remeis observatory and MIT
(\url{http://www.sternwarte.uni-erlangen.de/isis/}).

\Change{
This research employs a list of Chandra datasets, obtained by the
Chandra X-ray Observatory, contained in~\dataset[DOI:10.25574/cdc.457]{https://doi.org/10.25574/cdc.457}.
}

\end{acknowledgments}





\facilities{CXO (HETG), NuSTAR}

\software{
  \textsc{CIAO} \citep{CIAO:2006},
  \textsc{ISIS} \citep{houck2000}, 
  \textsc{HEASOFT} (\url{https://heasarc.gsfc.nasa.gov/lheasoft/})
}

\bibliography{v750}{}
\bibliographystyle{aasjournalv7}

\end{document}

%% file: v750_tbl_starparam.tex

\begin{deluxetable}{ccc}
  \small
  \tablecaption{\vsfz Parameters\label{tbl:starparam}}
  \tablewidth{0pt}
  \tablehead{
    \colhead{Property}& 
    \colhead{Value}&
    \colhead{Reference}
  }
  \startdata
  HD&
  157832&
  \\
  Spectral Type&
  B1.5 Ve&
  1
  \\
  Primary Mass&
  $11\msun$&
  1
  \\
  Secondary Mass&
  $0.9\pm0.1\msun$&
  4
  \\
  &
  $0.7-0.8\msun$&
  2
  \\
  Distance&
  $972$~pc&
  3
  \\
  $E(B-V)$&
  $0.24$&
  1
  \\
  %
  \Change{Projected Rotational Velocity}&
  $220\kms$&
  1
  \\
  %
  \Change{Primary Velocity Amplitude}&
  $6.25\kms$&
  2
  \\
  %
  \Change{Orbital Inclination}&
  $60$--$80^\circ$&
  2
  \\
  Orbital Period&
  $95.23\,$d&
  2
  \\
  Epoch&
  2458566.30 HJD&
  2
  \\
  \enddata
\tablecomments{\footnotesize
  1 -- \citet{lopes:motch:2011};
  2 -- \citet{wang:gies:al:2023};
  3 -- \citet{bailerjones:al:2021};
  4 -- this work.  The Epoch is the time of inferior conjunction of the secondary (that is, the secondary is nearest to the observer).
}
\end{deluxetable}


%% file: v750_tbl_obs.tex
%
\begin{deluxetable}{cccc}
\tablecaption{\vsfz Observational Information\label{tbl:obsinfo}}
\tablewidth{0pt}
\tablehead{
\colhead{ObsID}& 
\colhead{DATE-OBS}&
\colhead{Exposure} & 
\colhead{$\phi_\mathrm{orb}$} \\
 &
 \colhead{[start date]}&
 \colhead{[$\ks$]}&
}
\startdata
\dataset[27032]{\doi{10.25574/27032}}& 2023-02-01T13:10:35&    13.60&   0.81\\
\dataset[27689]{\doi{10.25574/27689}}& 2023-02-03T19:59:47&    14.58&   0.84\\
\dataset[27031]{\doi{10.25574/27031}}& 2023-09-23T23:37:22&    19.70&   0.28\\
\dataset[27033]{\doi{10.25574/27033}}& 2024-01-29T00:36:40&     9.44&   0.61\\
\dataset[29232]{\doi{10.25574/29232}}& 2024-01-29T08:11:10&    11.59&   0.61\\
\dataset[29233]{\doi{10.25574/29233}}& 2024-01-29T16:17:47&    12.30&   0.62\\
\dataset[29234]{\doi{10.25574/29234}}& 2024-01-30T18:27:24&     9.24&   0.63\\
30901001002& 2024-02-03T19:46:09&    95.21&   0.67\\                
\dataset[29374]{\doi{10.25574/29374}}& 2024-04-20T16:09:05&    25.60&   0.48\\
\dataset[26494]{\doi{10.25574/26494}}& 2024-04-21T11:52:03&    14.87&   0.49\\
\dataset[27034]{\doi{10.25574/27034}}& 2024-06-12T18:24:53&    17.49&   0.04\\
\dataset[29448]{\doi{10.25574/29448}}& 2024-06-15T23:59:11&    17.83&   0.07\\
\dataset[30463]{\doi{10.25574/30463}}& 2024-10-03T23:28:11&    17.34&   0.22\\
\dataset[27035]{\doi{10.25574/27035}}& 2024-10-05T14:01:49&    13.78&   0.24\\
\enddata

\tablecomments{\raggedright The orbital phase is defined by the ephemeris of
  \citet{wang:gies:al:2023}.
  The 5-digit ObsIDs are from \chan, while the 11-digit ObsID refers
  to \nustar.
}
\end{deluxetable}

%% file: v750_tbl_fit_params.tex
\begin{deluxetable*}{ccccc}
\tablecaption{Model Fit Results}\label{tbl:plasmafit}
\tablewidth{0pt}
\tablehead{
\colhead{Parameter}& 
\colhead{Value}& 
\colhead{Error}&
\colhead{Unit}&
\colhead{Comment}
}
\startdata
%
$norm$&
$1.06$&
$0.06$&
$10^{-16}\mdot\,\mathrm\pc^{-2} $&
\\
$\dot{M}$&
$1.0$&
$0.1$&
$10^{-10}\mdot$&
Derived from the $norm$ and distance
\\
%
($\gamma$)&
$0$&
&
&
Cooling flow exponent; (0 means adiabatic)
\\
%
($\log T_\mathrm{min}$)&
$6.05$&
&
dex [K]&
Minimum emission measure temperature
\\
%
$\log T_\mathrm{max}$&
$8.41$&
$0.02$&
dex [K]&
Maximum emission measure temperature
\\
%
($v_\mathrm{turb}$) &
$600$&
&
$\kms$& 
Excess required over thermal broadening
\\
%
($\Delta v$) &
$0$&
&
&
Doppler shift
\\
%
$A$(Mg)&
$1.2$&
$0.3$&
&
Abundance factors relative to Solar
\\
%
$A$(Si)&
$1.3$&
$0.2$&
&
\\
%
$A$(S) &
$1.3$&
$0.4$&
&
\\
%
$A$(Fe)&
$1.3$&
$0.1$&
&
\\[2mm]
%
($N_\mathrm{H}(ISM)$)&
$0.20$&
&
$10^{22}\cmmtwo$&
Foreground interstellar absorption \citep{lopes:motch:2011}
\\
%
$N_\mathrm{H}$&
$0.40$&
$0.02$&
$10^{22}\cmmtwo$& 
System intrinsic absorption
\\[2mm]
%
$f_\mathrm{x}$&
$3.5$&
$0.04$&
$10^{-12}\eflux$&
Flux $0.3$--$20\kev$ (as observed with absorption)
\\
%
$L_\mathrm{x}$&
$4.1$&
$0.1$&
$10^{32}\lum$&
Luminosity $0.3$--$20\kev$ (only ISM absorption removed)
\\[2mm]
%
$f({\mathrm{Mg\,XI}})$&
$1.3$&
$0.6$&
$10^{-6}\pflux$&
He-like ion, resonance line
\\
%
$f(\mathrm{Mg\,XII})$&
$1.0$&
$0.4$&
$10^{-6}\pflux$& 
H-like ion, Ly-$\alpha$-like transition
\\
%
$f(\mathrm{Si\, XIV})$&
$3.5$&
$0.6$&
$10^{-6}\pflux$& 
H-like ion, Ly-$\alpha$-like transition
\\
%
$f(\mathrm{S\,XVI})$&
$1.4$&
$0.6$&
$10^{-6}\pflux$&
H-like ion, Ly-$\alpha$-like transition
\\
%
$f(\mathrm{Fe K})$&
$1.6$&
$0.5$&
$10^{-6}\pflux$&
Fluorescent emission, joint fit
\\ 
%
$f(\mathrm{Fe K})$&
$4.0$&
$0.9$&
$10^{-6}\pflux$&
Fluorescent emission, \nustar only
\\ 
%
$f({\mathrm{Fe\,XXV}})$&
$5.6$&
$0.4$&
$10^{-6}\pflux$&
He-like resonance line
\\ 
%
$f({\mathrm{Fe\,XXVI}})$&
$2.7$&
$0.6$&
$10^{-6}\pflux$&
H-like ion, Ly-$\alpha$-like transition
\\
%
\enddata
\tablecomments{\footnotesize Parameters in ``()'' were frozen.  Relative
  abundances are referenced to \citet{asplund:grevesse:al:2009}. \Change{The
  ``Error'' column gives the one standard deviation uncertainties.
  $\dot M$ and $L_\mathrm{x}$ assumed a distance of $972\pc$
  \citep{bailerjones:al:2021} and removed only the ISM component of
  the absorption.  Flux and luminosity are integrated from the model
  over the $0.3$--$20\kev$ band. The integrated flux is as-observed
  (uncorrected for any absorption). Line fluxes are from Gaussian fits
  to each feature, which were the \ChangeB{H-like Ly-$\alpha$} or the
  He-like resonance lines, without correction for absorptions.}
}
\end{deluxetable*}